\documentclass[twocolumn,9pt]{article} 

\usepackage[square,numbers,sort&compress,comma]{natbib}

\usepackage{amsmath}
\usepackage{amssymb}
\usepackage{caption}
\usepackage{graphicx}
\usepackage{latexsym}
\usepackage{times}
\usepackage[pagewise]{lineno}
\usepackage{hyperref}
\usepackage[version=4]{mhchem}
\usepackage{siunitx}
\usepackage{booktabs}
\usepackage{cleveref}
\crefname{figure}{Fig.}{Figs.} 
\Crefname{figure}{Figure}{Figures} 
\crefname{equation}{Eq.}{Eqs.} 
\topmargin - 12pt 
\renewenvironment{abstract}%
              {
               \small
               {\bfseries \abstractname}
               \par
               \vspace{10pt}
              }

\renewcommand\abstractname{Abstract}

\newcommand{\nomenclature}
              [1]
              {
               \bgroup
               \flushleft
               \small\bf
               #1
               \par
               \egroup
              }

\renewcommand{\section}
              [1]
              {
               \bgroup
               \flushleft
               \small\bf
               \refstepcounter{section}
               \arabic{section}. #1
               \par
               \egroup
              }

\renewcommand{\subsection}
              [1]
              {
               \bgroup
               \flushleft
               \small\em
               \refstepcounter{subsection}
               \arabic{section}.
               \arabic{subsection}. #1
               \par
               \egroup
              }

\renewcommand{\subsubsection}
              [1]
              {
               \bgroup
               \flushleft
               \small\em
               \refstepcounter{subsubsection}
               \arabic{section}.
               \arabic{subsection}.
               \arabic{subsubsection}. #1
               \par
               \egroup
              }

  \newcommand{\acknowledgement}
              [1]
              {
               \bgroup
               \flushleft
               \small\bf
               #1
               \par
               \egroup
              }

  \newcommand{\sectionbib}
              [1]
              {
               \bgroup
               \flushleft
               \small\bf
               #1
               \par
               \egroup
              }

\begin{document}



\small
\baselineskip 10pt

\setcounter{page}{1}
\title{\LARGE \bf Flow configuration and pressure effects on turbulent premixed hydrogen jet flames}

\author{{\large T.~L.~Howarth$^{a,b,*}$, T.~Lehmann$^{b}$, M.~Gauding$^{b}$, H.~Pitsch$^{b}$}\\[10pt]
        {\footnotesize \em $^a$Department of Aeronautical and Automotive Engineering, Loughborough University, Loughborough LE11 3TU, UK}\\[-5pt]
        {\footnotesize \em $^b$Institute for Combustion Technology, RWTH Aachen University, 52056 Aachen, Germany}}

\date{}  

\twocolumn[\begin{@twocolumnfalse}
\maketitle
\rule{\textwidth}{0.5pt}
\vspace{-5pt}

\begin{abstract} 
Turbulent lean premixed hydrogen jet flames are simulated using direct numerical simulation employing detailed chemistry in both slot and round configurations at various pressures. All cases are simulated at a constant jet Reynolds number ($Re_j = 10\,000$) and a fixed ratio of characteristic length scales. While normalised macroscopic quantities (e.g., flame length, turbulent flame speed) appear comparable across configurations, fundamental discrepancies are observed that originate from the coupling of large- and small-scale effects. Mean local reactivity ($I_0$) decays monotonically downstream, driven by a decreasing Karlovitz number ($Ka^{*}$); however, this decay is modulated by geometry, with round jets exhibiting a faster decline due to mean negative curvature. Pressure is identified as a critical small-scale driver, fundamentally altering flame propagation by increasing the sensitivity of displacement speed to local curvature. At elevated pressures, this sensitivity induces higher flame stretch and accelerates wrinkling near the nozzle, which compounds with geometry-dependent effects, such as the slower decay of mean strain in slot configurations. 
\end{abstract}

\vspace{10pt}

{\bf Novelty and significance statement}

\vspace{10pt}

The novelty of this research is the first systematic comparison of lean premixed hydrogen jet flames in slot and round configurations across varying pressures. While key trends identified in homogeneous isotropic turbulence serve as a robust baseline for these flows, the study reveals configuration- and pressure-specific discrepancies. This is significant because it links findings from idealised turbulence studies and practical, anisotropic combustion environments. Furthermore, the work identifies a previously unreported regime at elevated pressures where flame propagation transitions from surface-area destruction to net surface-area generation due to intensified thermodiffusive effects. 


\vspace{5pt}
\parbox{1.0\textwidth}{\footnotesize {\em Keywords:} Turbulent premixed jet flame; Direct numerical simulation; Lean premixed hydrogen; High-pressure combustion; Differential diffusion}
\rule{\textwidth}{0.5pt}
*Corresponding author.
\vspace{5pt}
\end{@twocolumnfalse}] 

\section{Introduction\label{sec:introduction}} \addvspace{10pt}

Turbulent jet flames are a common configuration in industrial settings, and with the ongoing aim of net-zero emissions, the usage of hydrogen (\ce{H2}) proves to be an attractive alternative for such applications. To mitigate the production of nitrogen oxide (\ce{NO}) and reduce the burning velocity of the flame, it is preferable to burn \ce{H2} in a lean premixed mode, which comes with technical challenges associated with potentially strong differential diffusion effects due to the high mobility of \ce{H2}~\cite{pitsch2024transition}. One of the main challenges is the prediction of enhanced burning velocities due to thermodiffusive effects~\cite{day2009turbulence,aspden2011turbulence,howarth2022empirical,howarth2023thermodiffusively}, which, in the context of jet flames, drastically reduces flame length~\cite{berger2022synergistic,berger2024effects}, increasing the propensity for flashback. Pressure is known to strongly affect thermodiffusive instability in lean premixed \ce{H2} flames~\cite{howarth2023thermodiffusively,rieth2023effect,berger2022intrinsic}, with a strong response due to the enhancement of Zeldovich ($Ze$) number by increased importance of radical recombination reactions~\cite{law2006propagation}. Turbulence is also known to strongly interact with thermodiffusive effects, with the normalised local reactivity (or stretch factor) $I_{0}$ increasing with Karlovitz number~\cite{day2009turbulence,howarth2023thermodiffusively,berger2022synergistic}. In a jet setup, two configurations are common, namely slot and round geometries for the inflow. In an experimental setting, premixed \ce{H2} jet flames have been studied primarily in a round jet configuration~\cite{wu1990turbulent}. In direct numerical simulation (DNS) studies, slots are generally preferred~\cite{berger2022synergistic,macart2019evolution}, with round configurations only appearing in works that perform direct comparisons with experiments~\cite{wang2016turbulence,wang2017direct,russell2025turbulence}. In an incompressible jet, these two flows have different self-similarity properties and the decay of turbulence behaves differently~\cite{pope2000turbulent}. It would therefore be expected that the flame may behave differently between these configurations due to the variation of turbulence-flame interactions in the streamwise direction. 

This work aims to understand both the effect of inflow configuration choice (slot or round inlets) and pressure on the turbulent burning velocity through the enhancement or suppression of wrinkling and normalised local reactivity resulting from underlying differential diffusion or flow field effects at both large and small scales. 

Section \ref{sec:dns} outlines the solver and numerical configuration. Analysis of the database is discussed in section \ref{sec:results}, with a focus on turbulent burning velocity and normalised local reactivity (section \ref{subsec:st}), and flame surface area generation (section \ref{subsec:area}). The findings are discussed and conclusions are provided in section \ref{sec:conclusions}.

\section{Direct numerical simulation database\label{sec:dns}} \addvspace{10pt}

In this section, the solver and setup for the simulations are briefly discussed.

\subsection{Numerical solver\label{subsec:solver}} \addvspace{10pt}

Three-dimensional direct numerical simulations (DNS) were performed using the PeleLMeX solver~\cite{esclapez2023pelelmex}, which solves the low-Mach number reacting Navier-Stokes equations for an ideal gas mixture. The transport properties utilise a mixture-averaged diffusion model, with the Soret effect included following \cite{howarth2024thermal}. The governing equations are discretised using a second-order Godunov finite-volume method, where advection, diffusion, and reaction terms are integrated via a spectral deferred correction approach~\cite{nonaka2018conservative}. A density-weighted projection method enforces the divergence constraint, satisfying the equation of state~\cite{almgren1998conservative, day2000numerical}. Time integration is controlled by an advective CFL condition, with diffusive and reactive source terms treated implicitly. This scheme is implemented within the AMReX block-structured adaptive mesh refinement (AMR) framework~\cite{zhang2019amrex}. Transport coefficients, thermodynamic relationships, and chemical kinetics were taken from a detailed chemical mechanism suitable for high-pressure hydrogen combustion~\cite{burke2012comprehensive}.

\subsection{Simulation configuration\label{subsec:config}} \addvspace{10pt}
The simulations were set up with slot or round inflows on the low-$z$ face, with a precursor simulation performed in a periodic channel or pipe configuration sampled temporally to generate turbulent inflows with a Reynolds number of 10\,000. Outside of the inflow, a laminar coflow is used, set at 15\% of the bulk velocity of the jet ($u_{j}$). In all simulations, outflow boundaries are imposed on the high-$z$ face and both $x$ directions. In the slot configuration, the $y$ boundaries are set as periodic, and as outflows in the round configuration. The domain sizes ($L_{x},L_{y},L_{z}$) in all cases are normalised by the jet width $H$ (or diameter $d_{j}$), and for each case, there are 17 characteristic flame thicknesses ($l_{F}^{*}$) across the inlet. $l_{F}^{*} = l_{F}\exp(-0.06\omega_{2})$, where $l_{F}$ is the unstretched laminar flame thickness and $\omega_{2}$ is the instability parameter used for modelling the thermodiffusive response~\cite{howarth2022empirical,howarth2023thermodiffusively}. $u_{j}$ is then scaled to match the jet Reynolds number across each case. In all cases, the streamwise distance is set at $16H$ (or $16d_{j}$). All details can be found in table \ref{tab:sim_params}, where $p_{0}$ is the thermodynamic pressure, $s_{L}$ is the unstretched laminar flame speed and $s_{L}^{*} = s_{L}\exp(0.08\omega_{2})$ is the characteristic flame speed. For the round cases, a square base of 4 jet diameters is used, whereas in the slot case, the cross-stream ($x$) size is set at 12 jet widths, with 4 jet widths in the periodic ($y$) direction. The usage of $^{*}$ (or freely-propagating) quantities, rather than unstretched quantities, is both important and significant, as it results in slightly smaller jets for the atmospheric cases, but much smaller jets for the elevated pressure cases. 

\begin{table}[h]
    \centering
    \caption{Simulation parameters; any adjustments made for the round cases are denoted in parentheses.}
    \label{tab:sim_params}
    \footnotesize {
    \begin{tabular}{l c c c }
        \toprule
        Case & S(R)1 & S(R)5 & S(R)10 \\
        \midrule
        $p_{0}$ (atm) & 1 & 5 & 10  \\
        $\omega_2$ & 5.94 & 19.9 & 29.8 \\        
        $l_F$ (\si{\micro\meter}) & 673 & 369 & 608 \\
        $s_L$ (\si{\centi\meter\per\second}) & 20.1 & 5.17 & 1.52 \\        
        $l_F^{*}$ (\si{\micro\meter}) & 471 & 112 & 102 \\
        $s_L^{*}$ (\si{\centi\meter\per\second}) & 32.3 & 25.4 & 16.5 \\        
        $u_{j}$ (\si{\meter\per\second}) & 24.0 & 19.2 & 10.6 \\
        $H,d_{j}$ (\si{\milli\meter}) & 8 & 2 & 1.8 \\
        $L_{x}$ (\si{\milli\meter}) & 96 (32) & 24 (8) & 21.6 (7.2) \\
        $L_{y}$ (\si{\milli\meter}) & 32 & 8 & 7.2 \\
        $L_{z}$ (\si{\milli\meter}) & 128 & 32 & 28.8 \\
        $\Delta x$ (\si{\micro\meter}) & 31.3 & 7.81 & 7.03 \\
        \bottomrule
    \end{tabular}
    }
\end{table}

\section{Results\label{sec:results}} \addvspace{10pt}
In this section, the turbulent burning velocity is firstly examined, along with the enhanced normalised reactivity in section \ref{subsec:st}, followed by an analysis of the flame surface area generation mechanism in section \ref{subsec:area}.

\subsection{Turbulent burning velocity and normalised local reactivity\label{subsec:st}} \addvspace{10pt}

\begin{figure}
    \centering
    \includegraphics[trim=7cm 0cm 7cm 8cm, clip,width=0.3\linewidth]{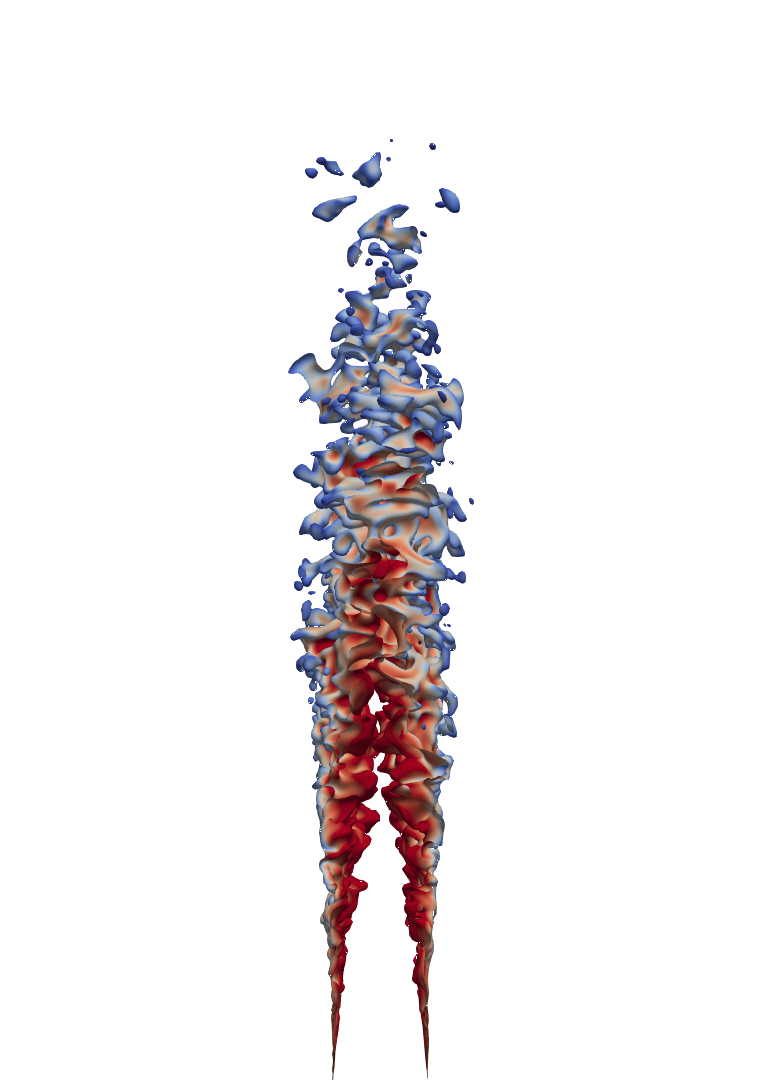}
    \includegraphics[trim=7cm 0cm 7cm 8cm,  clip,width=0.3\linewidth]{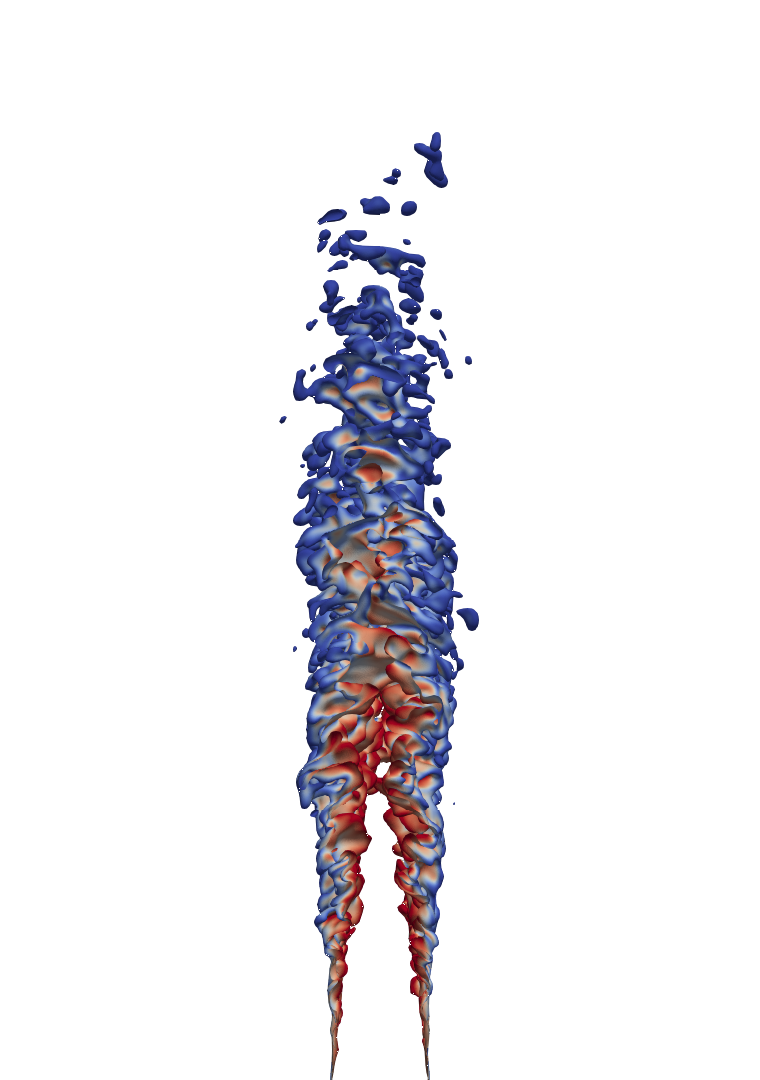}
    \includegraphics[trim=7cm 0cm 7cm 8cm,  clip,width=0.3\linewidth]{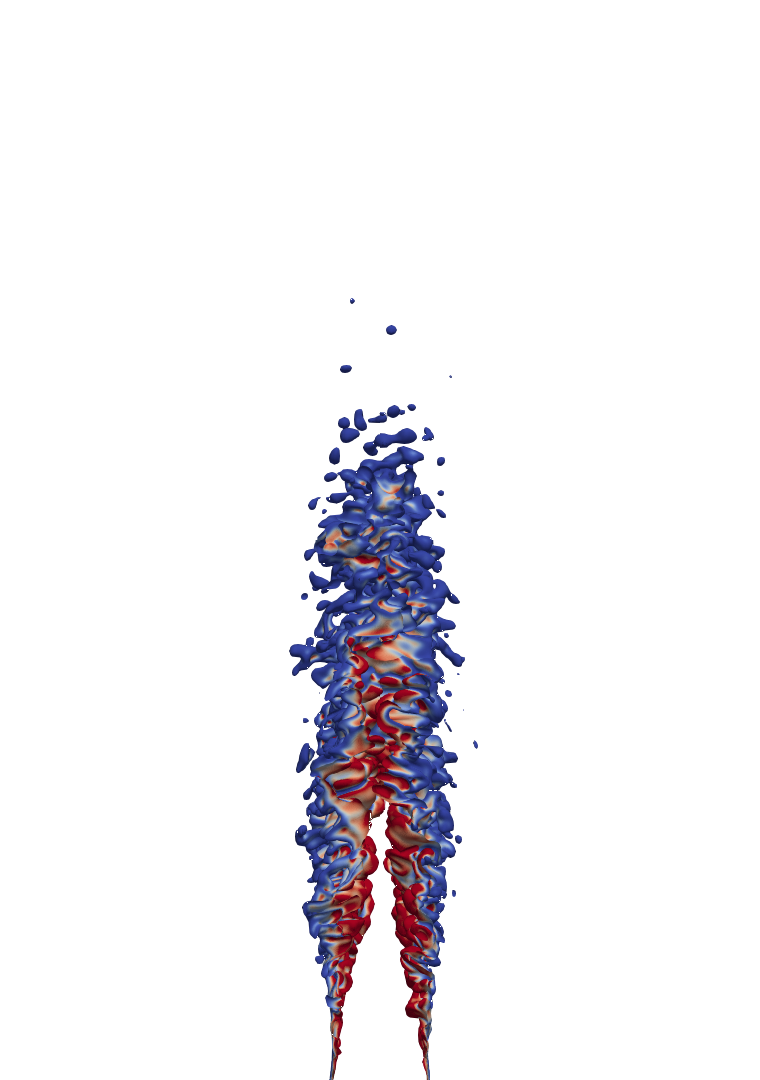}

    \includegraphics[trim=7cm 0cm 7cm 8cm,  clip,width=0.3\linewidth]{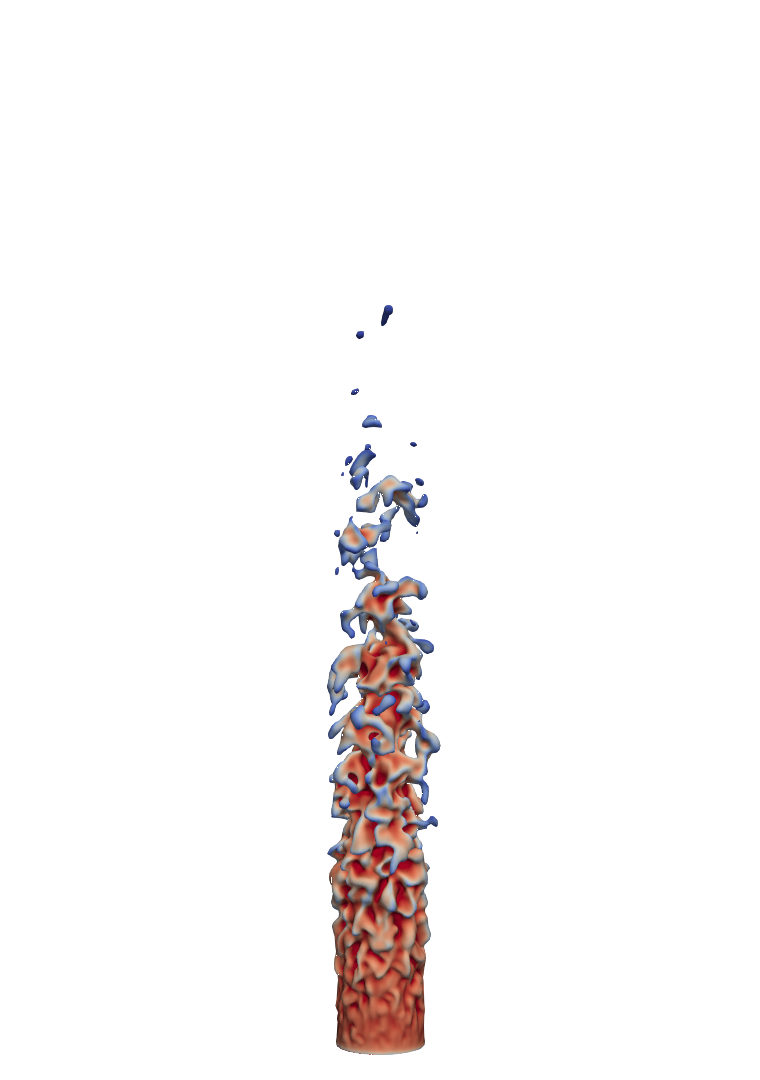}
    \includegraphics[trim=7cm 0cm 7cm 8cm, clip,width=0.3\linewidth]{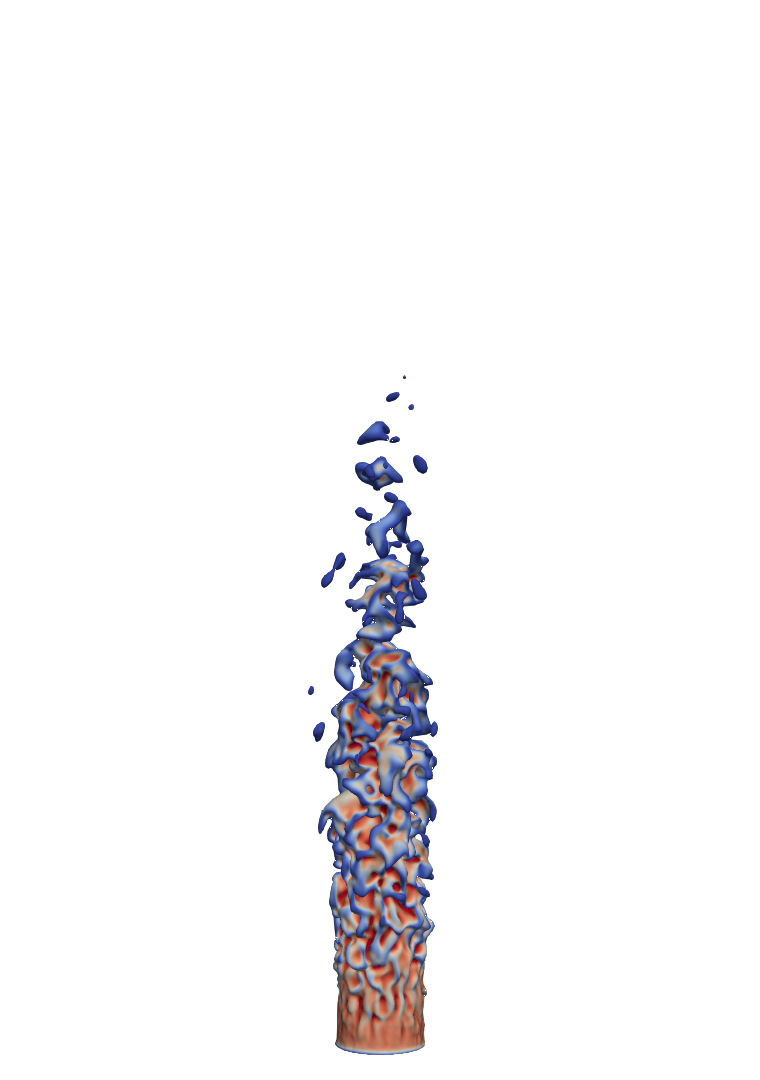}
    \includegraphics[trim=7cm 0cm 7cm 8cm, clip,width=0.3\linewidth]{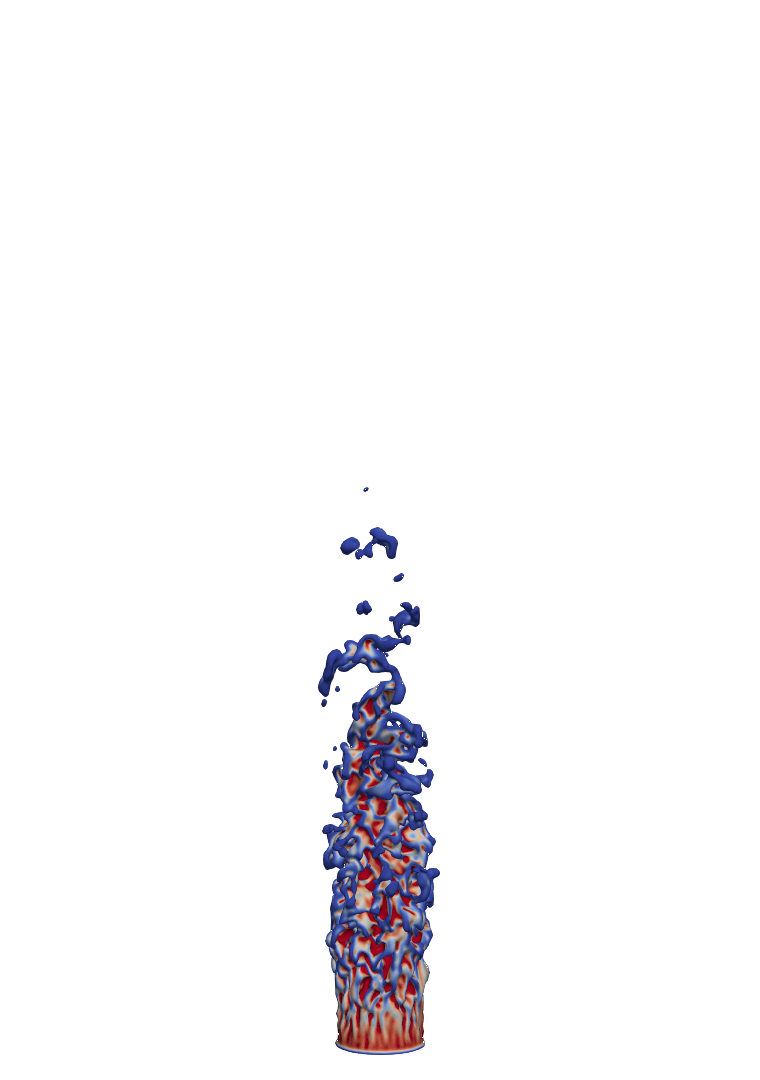}
    \caption{Isosurfaces of each case at $c_{F} = 0.9$, coloured by heat release rate. Top row: slot cases, bottom row: round cases, left column: 1atm, middle column: 5atm, right column: 10atm. Each case normalised in size by $H$ or $d_{j}$.}
    \label{fig:surfaces}
\end{figure}

Sample isosurfaces of the flames, defined at an isosurface determined by the fuel-based progress variable
\begin{equation}
    c_{F} = 1 - \frac{Y_{\ce{H2}}}{Y_{\ce{H2},u}}
\end{equation}
with $c_{F} = 0.9$ and coloured by local heat release, can be seen in \cref{fig:surfaces}. In both slot and round cases, there is significant variation in reactivity across the flame, and this variation is enhanced by the increase in pressure. Visually, despite the usage of normalised length scales, the flame also decreases in height as the pressure increases. 

\begin{figure}
    \centering
    \includegraphics[width=\linewidth]{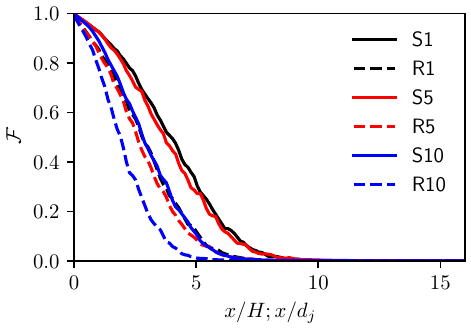}
    \caption{Fuel flux as a function of streamwise distance for all cases. Slot cases are denoted with solid lines, round cases with dotted lines. Pressures of 1, 5 and 10atm are shown with black, red and blue lines, respectively.}
    \label{fig:ff}
\end{figure}

The flame length can be determined by examining the normalised fuel flux, given by
\begin{equation}
    \mathcal{F} = \frac{1}{\rho_{u}Y_{\ce{H2},u}A_{j}u_{j}}\int_{0}^{L_{x}}\int_{0}^{L_{y}}\rho Y_{\ce{H2}}u_{z}\, dx dy
\end{equation}
where $\rho$ is the density, $Y_{\ce{H2}}$ is the mass fraction of \ce{H2}, and $A_{j}$ is the area of the jet. \cref{fig:ff} shows the fuel flux as a function of streamwise distance. The longest cases occur for slot burners at lower pressure, whereas the shortest occur for round burners at higher pressure. 

\begin{figure}
    \centering    
    \includegraphics[width=\linewidth]{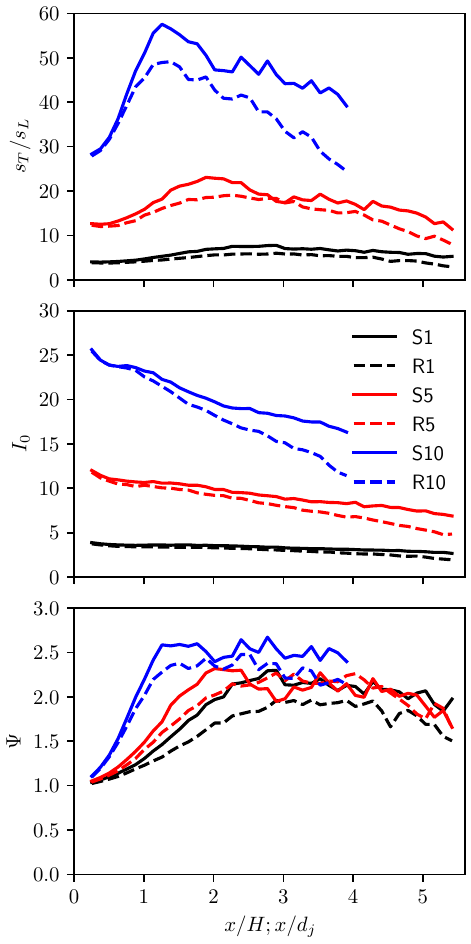} 
    \caption{Normalised turbulent burning velocity $s_{T}/s_{L}$ (top), mean local reactivity $I_{0}$ (middle) and flame wrinkling $\Psi$ (bottom) as a function of streamwise distance.}
    \label{fig:st}
\end{figure}

The flame length is a direct result of the turbulent burning velocity and the area of the inlet. The former can be decomposed into flame wrinkling and normalised local reactivity:
\begin{equation}
    s_{T}/s_{L} = I_{0}\Psi, \Psi  = A_{T}/A_{0}
\end{equation}
where $s_{T}$ is the turbulent burning velocity, $I_{0}$ is normalised reactivity, $\Psi$ is the flame wrinkling, $A_{T}$ is the instantaneous flame surface area and $A_{0}$ is the flame surface area of the mean field, both of which are evaluated using the fuel field, and evaluated at $c_{F}=0.9$ ($c_{F} = 1-Y_{\ce{H2}}/Y_{\ce{H2},u}$). Each of these quantities is plotted along the streamwise direction for each case in \cref{fig:st}. In all cases, there is an increase of $s_{T}$ near the nozzle before reaching a local maximum and then decreasing towards the flame tip. By comparing $I_{0}$ and $\Psi$, this can be seen to be due to the sharp increase of wrinkling up to some roughly constant value, which is marginally higher for the higher pressure and slot cases. The $I_{0}$ on the other hand, monotonically decreases. This has been noted before in other studies~\cite{berger2022synergistic,howarth2025structure}, where it was attributed to a combined effect of a decaying effect of $Ka^{*}$, and the mean curvature of the round jet. Here, a direct comparison is made between the slot and round jet to concretely demonstrate this. Although the slot jets exhibit the highest burning velocities (i.e.~burn more fuel per second), the round jets are shorter due to the considerably smaller inlet area, which results in a lower mass flow rate.

\begin{figure}
    \centering
    \includegraphics[width=\linewidth]{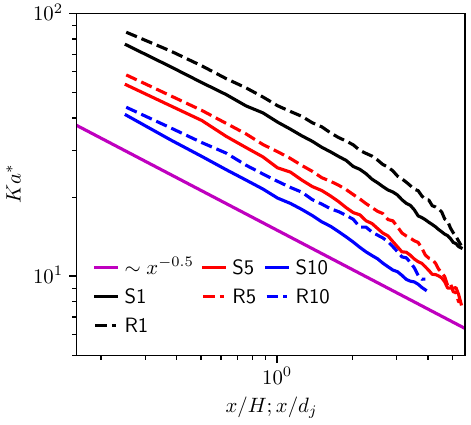}
    \caption{Karlovitz number $Ka^{*}$ as a function of the normalised streamwise distance. All cases exhibit a $\sim x^{0.5}$ decay, shown by the magenta line.}
    \label{fig:Ka}
\end{figure}

\cref{fig:Ka} shows the evolution of $Ka^{*}$ as a function of streamwise distance, which is defined as
\begin{equation}
    Ka^{*} = \sqrt{\frac{\langle \varepsilon\rangle_{s|z}^{c_{F} = 0.1}}{\varepsilon_{F}^{*}}},
    \, \varepsilon = \widetilde{\tau_{ij}^{\prime\prime}\frac{\partial u_{i}^{\prime\prime}}{\partial x_{j}}}, \varepsilon_{F}^{*} = \frac{s_{L}^{*3}}{l_{F}^{*}},
\end{equation}
where $\tau_{ij}$ is the viscous stress tensor and the $\tilde{.}$ operation represents a Favre-average. Note that for the Karlovitz numbers ,quantities, and their respective fluctuations, are computed directly on the surface~\cite{howarth2025structure}. It can be seen that $Ka^{*}$ decays with an approximate power law of $x^{-0.5}$ ($\varepsilon \sim x^{-1}$) regardless of the flow configuration or pressure level. Note that this is not consistent with typical self-similarity theory for incompressible jets, but is an empirical observation in the near-field, sampled close to the flame, rather than along the centreline. From these figures, it is also clear that $Ka^{*}$ seems to be largely insensitive to the flow configuration. $Ka^{*}$ quantifies turbulence-flame interactions at the inertial subrange, where the effects of larger-scale structures are less relevant. Pressure appears to slightly reduce the $Ka^{*}$. Although the ratio between the length scales ($d_{j}/l_{F}^{*}$) has been maintained, the ratio between velocity scales ($u_{j}/s_{L}^{*}$) has not. This results from a reduction in the flame Reynolds numbers ($Re_{F} = s_{L}^{*}l_{F}^{*}/\nu$) with increasing pressure, reducing the velocity ratios; this in turn reduces the $Ka^{*}$.

\begin{figure}
    \centering
    \includegraphics[width=\linewidth]{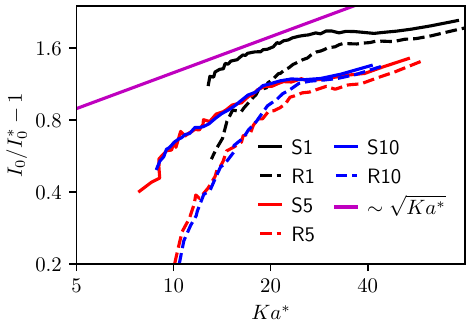}
    \caption{Turbulent enhancement of $I_{0}$ ($I_{0}/I_{0}^{*} -1$) plotted as a function of $Ka^{*}$. Note that the inlet is positioned on the right, and the flame tip on the left.}
    \label{fig:I0Ka}
\end{figure}

\Cref{fig:I0Ka} shows the turbulent enhancement of $I_{0}$ from each flame's laminar value ($s_{L}^{*} = I_{0}^{*}s_{L}$), $I_{0}/I_{0}^{*} - 1$. In homogeneous isotropic turbulence (HIT), it was found that for all conditions, this enhancement scales with $\sqrt{Ka^{*}}$~\cite{howarth2023thermodiffusively}. Here, slot cases match a $\sqrt{Ka}$ scaling in the developed turbulence region, i.e.~the region in the centre of the graph, away from the nozzle on the right-hand side, and also away from the flame tip, where there is considerable flame-flame interaction and negative mean curvatures. Additionally, the constant of proportionality decreases with increasing pressure, also consistent with findings from HIT. 

However, as seen in~\cite{howarth2025structure}, the decay in the round jets is faster. In HIT, it was observed that~\cite{howarth2023thermodiffusively}
\begin{equation}
    \frac{s_{loc}}{s_{L}} = I_{0} - \mathcal{M}\kappa l_{F}, \quad \mathcal{M} = -2
\end{equation}
where $s_{loc}$ is the local consumption-based flame speed and $\kappa$ is the mean curvature. When averaged over the slot and round configuration, a model can be constructed
\begin{equation}\label{eq:kmodel}
\begin{split}
     \frac{I_{0}^{r}}{I_{0}^{s}} &= \frac{I_{0} + 2\langle\kappa\rangle l_{F}}{I_{0}} \\ &= 1 +2 \langle\kappa\rangle l_{s},\,l_{s} = \frac{l_{F}}{I_{0}^{s}}
     \end{split}
\end{equation}
where $I_{0}^{r/s}$ are the $I_{0}$ in the round and slot configuration, respectively. 

\begin{figure}
    \centering
    \includegraphics[width=\linewidth]{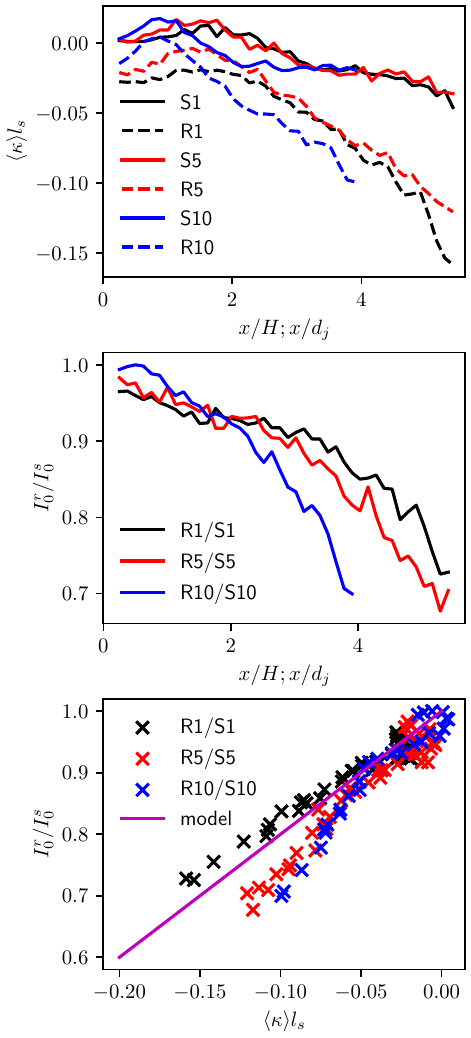}   
    \caption{Ratio of the $I_{0}$ between the round and slot cases as a function of streamwise distance (top) and normalised mean curvature (bottom).}
    \label{fig:Kamodel}
\end{figure}

The top panel of \cref{fig:Kamodel} illustrates the evolution of the normalised mean curvature $\langle\kappa\rangle l_s$ along the streamwise direction. While the slot cases (solid lines) remain relatively stable and close to zero, the round cases (dashed lines) show a decrease towards the flame tip. This geometric evolution correlates directly with the middle panel, where the ratio $I_{0}^{r}/I_{0}^{s}$ drops significantly below unity as $x/L_x$ increases. The bottom panel demonstrates the scaling collapse of the data. By plotting the ratio directly against the mean curvature, all of the cases align with the model from \cref{eq:kmodel} (magenta). This alignment shows that the faster decay observed in round jets is fundamentally a curvature-driven phenomenon, allowing the discrepancy to be captured by a simple linear correction to the planar baseline.

\subsection{Flame surface area generation\label{subsec:area}} 
\begin{figure}
    \centering
    \includegraphics[width=\linewidth]{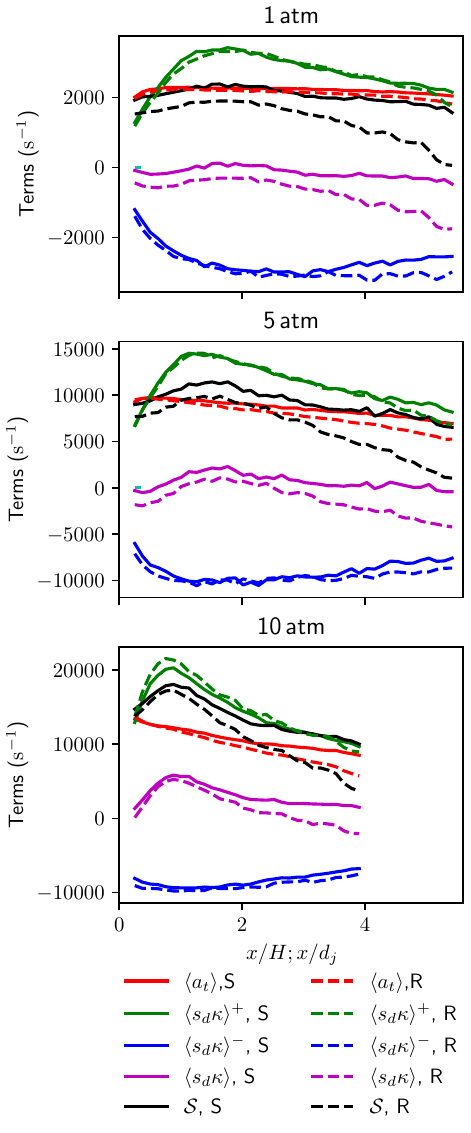}    
    \caption{Decomposed flame stretch statistics for the low (top), medium (middle) and high (bottom) pressure cases.}
    \label{fig:stretch}
\end{figure}

\addvspace{10pt}
To understand why the wrinkling trends are different between the different configurations and pressures, the flame stretch is computed, defined as~\cite{candel1990flame}
\begin{equation}
    \mathcal{S} = \frac{1}{A}\frac{d A}{dt} = \langle s_{d}\kappa\rangle + \langle a_{t}\rangle
\end{equation}
where $s_{d}$ is the flame displacement speed and $a_t$ is the tangential strain rate, defined as
\begin{align}\label{eq:sd}
    s_{d} &= \underbrace{-\frac{\dot\omega_{\ce{H2}}}{\rho\left|\nabla Y_{\ce{H2}}\right|}}_{s_{d,C}}
    \underbrace{- \frac{\nabla \cdot \left(\rho D_{\ce{H2}} \nabla X_{\ce{H2}}\right)}{\rho\left|\nabla Y_{\ce{H2}}\right|}}_{s_{d,D}} \\ a_{t} &= \nabla\cdot \mathbf{u} - \mathbf{nn:\nabla u}
\end{align}
These terms and their sum can be seen in \cref{fig:stretch}. The trend of the strain rate term is similar in each case, starting at some value (determined by the upstream boundary layer) and decaying with streamwise distance. However, unlike $\langle\varepsilon\rangle$, $\langle a_{t}\rangle$ exhibits sensitivity to the flow configuration, with faster decay observed in the round jet. Since $\langle a_{t}\rangle$ is evaluated on the overall field (rather than the fluctuating field like $\langle\varepsilon\rangle$), large-scale, mean flow field effects imprint themselves in this quantity. This difference can be associated with the faster decay of the mean velocity of the round jet compared to the slot jet, and leads to enhanced wrinkling in the slot jet.

The flame propagation term ($s_{d}\kappa$), on the other hand, strongly changes with pressure. At 1atm, this term solely destroys flame surface ($\langle s_{d}\kappa\rangle < 0$ everywhere), whereas it acts neutrally at 5atm ($\langle s_{d}\kappa\rangle  \approx 0$) and even generates additional area at 10atm ($\langle s_{d}\kappa\rangle > 0$). This explains the additional wrinkling, particularly near the nozzle, in the higher pressure cases. To determine the origin of the response at higher pressures, the flame propagation term is further decomposed into (following \cite{chu2023effects})
\begin{equation}
    \langle s_{d}\kappa\rangle = \underbrace{\langle s_{d}\kappa | \kappa > 0\rangle}_{\langle s_{d}\kappa \rangle^{+}} + \underbrace{\langle s_{d}\kappa | \kappa < 0\rangle}_{\langle s_{d}\kappa \rangle^{-}} 
\end{equation}
which is also shown in \cref{fig:stretch}. Flame surface destruction in negative curvature regions through flame propagation ($\langle s_{d}\kappa\rangle^{-}$) is relatively consistent throughout the flame. This is due to the relatively low reaction rates in these regions, leading to destruction primarily by diffusion~\cite{echekki1999analysis}. At low pressure, this term does not counteract the destruction in negatively curved regions. However, as pressure is increased, this term eventually outweighs the destruction, resulting in a net production of flame surface area near the nozzle from flame propagation alone. 

\begin{figure}
    \centering
    \includegraphics[width=\linewidth]{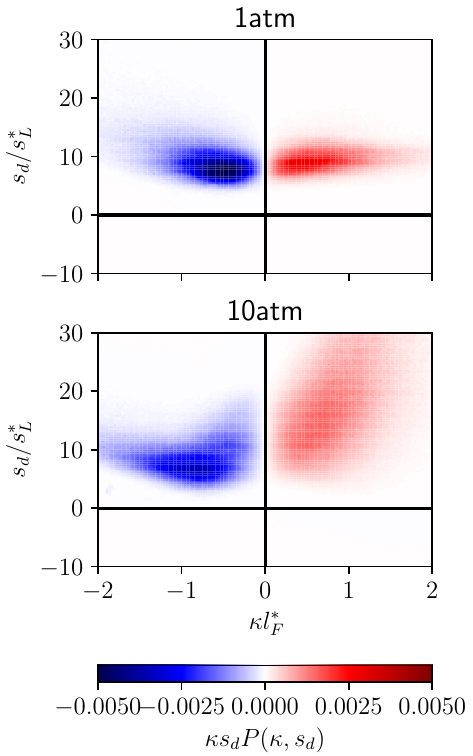}
    \caption{Premultiplied JPDF $\kappa s_{d}P(\kappa,s_{d})$ in the R1 (top) and R10 (bottom) cases. Blue areas of the heat map correspond to regions of the flame with flame surface area destruction, whereas red areas correspond to flame surface area generation.}
    \label{fig:sdbase}
\end{figure}

\begin{figure}
    \centering
    \includegraphics[width=\linewidth]{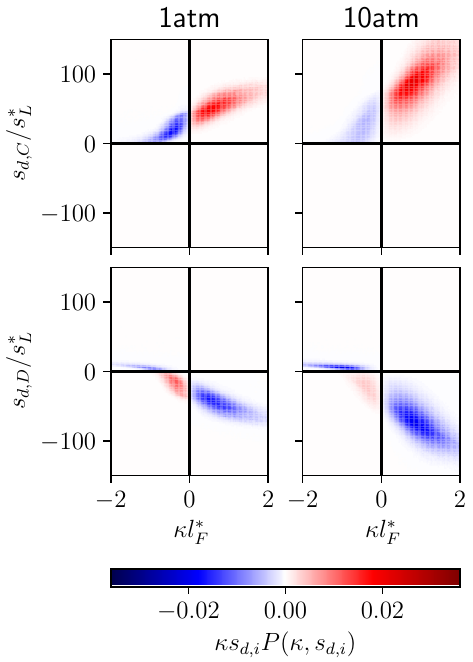}
    \caption{Premultiplied JPDFs decomposed into flame surface area changes due to chemistry $\kappa s_{d,C}P(\kappa,s_{d,C})$ and diffusion $\kappa s_{d,D}P(\kappa,s_{d,D})$.}
    \label{fig:sdsplit}
\end{figure}

To understand why this behaviour occurs, \cref{fig:sdbase} shows the premultiplied joint probability density function (JPDF) $\kappa s_{d} P(\kappa, s_{d})$ for the R1 and R10 cases. Integrating across this distribution yields the net area generation term $\langle s_{d} \kappa \rangle$, allowing for an examination of how different curvature regimes contribute to flame surface evolution. At atmospheric pressure, $s_{d}$ remains largely consistent across the curvature range. The distribution is primarily concentrated at negative curvatures, indicating that the net effect is surface area destruction. However, at 10 atm, $s_{d}$ becomes considerably more sensitive to curvature, exhibiting both a strong positive correlation and significantly increased scatter. To isolate the origin of this sensitivity, the displacement speed is decomposed into chemical ($s_{d,C}$) and diffusive ($s_{d,D}$) components (see \cref{eq:sd}) in \cref{fig:sdsplit}. At low pressure, these components exhibit comparable but opposite gradients ($s_{d,C} \sim -s_{d,D}$), resulting in a near-complete cancellation that maintains a quasi-steady flame structure. At 10\,atm, this balance is disrupted by the increase in the chemical term, which is greatly enhanced, resulting in a breakdown of the anti-correlation between the two terms. Crucially, this decoupling causes the net area generation to flip from negative to positive at high pressure, as the enhanced chemical response at positive curvatures overcomes the diffusive terms, leaving a noticeable, uncancelled, correlation with curvature.

\section{Discussion and conclusions\label{sec:conclusions}} \addvspace{10pt}

A database of three-dimensional direct numerical simulations of premixed hydrogen jet flames has been generated and analysed to uncover the effects of jet geometry and pressure on the flame. The normalised turbulent burning velocity ($s_{T}/s_{L}$) is maximised for the slot flames at the highest pressures considered (10 atm), and minimised in the round jet flame at the lowest pressure (1 atm). This variation is driven by a combined effect on mean local reactivity and flame wrinkling. In all cases, the turbulence dissipation rate $\varepsilon$ (and hence $Ka^{*}$) monotonically decreases in the streamwise direction, resulting in a decrease of $I_{0}$. In the slot configuration, the scaling approaches that found in homogeneous isotropic turbulence ($I_{0}/I_{0}^{*} - 1 \sim \sqrt{Ka}$), whereas the decay of $I_{0}$ is enhanced in the round configuration due to the negative and rapidly decreasing mean curvature. Comparing the ratio between the $I_{0}$ in the slot ($I_{0}^{s}$) and round ($I_{0}^{r}$) jets reveals a similar Markstein model to that in HIT, while flame wrinkling is enhanced in slot cases due to a slower decay of the mean strain rate.
As the system pressure increases, positively curved regions of the flame accelerate significantly faster than the mean flame propagation, resulting in rapid wrinkling near the nozzle. Decomposed analysis of the displacement speed reveals that this is driven by a fundamental shift in the balance between chemistry and diffusion. At low pressures, the displacement speed is relatively uniform across varying curvatures due to the approximate cancellation of chemical and diffusive terms. At higher pressures, however, the chemical response to positive curvature is significantly intensified, while the diffusive term fails to compensate. This leads to a marked increase in displacement speed and statistical scatter. Consequently, the net flame surface area evolution flips from destruction to generation at high pressure, as the chemistry-driven generation in positively curved regions outpaces the diffusion-driven destruction in negatively curved regions.

\acknowledgement{CRediT authorship contribution statement} \addvspace{10pt}

{\bf TLH}: Conceptualization, Methodology, Software, Formal analysis, Data curation, Writing - original draft. {\bf TL}: Conceptualization, Writing - review \& editing. {\bf MG}: Conceptualization, Writing - review \& editing. {\bf HP}: Conceptualization, Project administration, Funding acquisition, Writing - review \& editing. 

\acknowledgement{Declaration of competing interest} \addvspace{10pt}

The authors declare that they have no known competing financial interests or personal relationships that could have appeared to influence the work reported in this paper.

\acknowledgement{Acknowledgments} \addvspace{10pt}

TLH and HP acknowledge the financial support from Deutsche Forschungsgemeinschaft (DFG) within the project (ID: 516338899) IRTG 2983 Hy-Potential. TL, MG and HP acknowledge financial support from the European Research Council (ERC) Advanced Grant (HYDROGENATE, ID: 101054894). The authors gratefully acknowledge the Gauss Centre for Supercomputing e.V. (www.gauss-centre.eu) for funding this project by providing computing time on the GCS Supercomputers JUWELS at J\"ulich Supercomputing Centre (JSC) (ID: h2ex).

\footnotesize
\baselineskip 9pt

\clearpage
\thispagestyle{empty}
\bibliographystyle{proci}
\bibliography{PCI_LaTeX}


\newpage

\small
\baselineskip 10pt


\end{document}